\documentclass{article}%

\usepackage{amssymb}%
\usepackage{graphicx}

\begin{document}

\title{Comparative Study of Blockchain Development Platforms: Features and Applications}
\author{Collin Connors, Dilip Sarkar
\\The University of Miami}
\date{October 03, 2022}
\maketitle

\begin{abstract}
Many developers have ideas to create blockchain applications but do not know where to begin. Often these developers default to using the first blockchain development platform they discover, which may not be the best platform for their project. Over 8000 new blockchain-related projects are added to GitHub a year. The near-constant influx of new projects can make it difficult for developers to search through existing projects and platforms to find the best platform for their projects. We considered 65 blockchain development platforms for this work and provided a brief yet comprehensive summary of the 23 most popular platforms. Our aim for this work is to assist developers in selecting the most appropriate platform for their blockchain projects. 
\end{abstract}

\section{Introduction}
\label{sec:Introduction}
Satoshi Nakamoto, in the seminal paper Bitcoin: A peer-to-peer electronic cash system~\cite{NakamotoBitcoin} introduced the world to blockchain. While the system described in this paper is never explicitly called a blockchain, it has all of the properties of modern blockchains. Blockchains, as we have now come to know them, are tamper-evident and tamper-resistant digital ledgers implemented in a distributed fashion~\cite{NISTBlockchain}. This technology initially led to the rise of cryptocurrencies such as Bitcoin. 

However, as blockchains became more popular, developers began to expand the use cases of blockchains. This second generation of blockchains allowed developers to create smart contracts, code that can be run on a blockchain~\cite{SmartContractDef}. With smart contracts, developers could create decentralized applications governed by a blockchain. These decentralized applications are called dApps, which have become central in Web3 development. 

Web3 development depends on many ideas brought over from traditional web development, sometimes referred to as Web2 development. To create a traditional application in Web2 development, developers create and manage a front end and a back end layer. The front-end layer is responsible for displaying information to a user. This layer often relies on technology such as HTML, CSS, and JavaScript so a web browser can render a page. The back-end layer is responsible for managing the data through databases, APIs, and web hosting. Web2 developers use both the front-end layer and the back-end layer to create dynamic web applications. 

In Web3 development, a developer still has a front-end and a back-end layer in addition to a new smart contract layer. The smart contract layer allows Web3 developers to interact with a blockchain to create dApps. This new layer interacts with the front-end and back-end layers. However, the smart contract layer allows developers to create applications that do not have a centralized entity managing the data. 

In Web2 development, developers have many choices of development platforms for the front-end and back-end layers, such as various JavaScript frameworks or back-end scripting languages. In Web3 development, developers are presented with similar choices for the smart contract layer. Critically in order to create a dApp, a developer needs to select an appropriate blockchain development platform for their project. 
 
In Web2 development, one of new developers' most challenging choices is selecting the appropriate front-end and back-end technology stack for a project. The same problem exists in Web3 when selecting a blockchain development platform. There are many blockchain development platforms, with new ones being created yearly. This can make it difficult for developers to select the platform that best complements their projects. 

Our motivation for this work is to simplify the processes of selecting a blockchain development platform to make it easier for developers to get started on creating dapps. To do this, we have considered 65 blockchain development platforms and selected the 23 most popular platforms to analyze. With our analysis, we aim to highlight the differentiators between various blockchain development platforms so that developers can understand which platform will best synergize with their projects.

Previous surveys, while extensive, fail to provide everything a potential developer needs to know. These works only cover a few blockchain development platforms~\cite{valentaEtheriumHyperledgerCorda2017}, only cover blockchain development platforms for a specific industry~\cite{AgboHealthcareBlockchains2019}, only cover specific types of blockchain development platforms~\cite{QuasimBlockchainIoT2020}, \cite{PolgePermissionedBlockchainSurvey201}, \cite{IndustrySurvey_2021},  or do not provide enough details necessary to select the most appropriate platform \cite{TavaresSurveyOnBlockchainResearch2019}, \cite{DernaykaBlockchainPlatforms}. 

In previous work~\cite{BCCA_2022}, we performed a comparable study to the one presented here but only on 15 platforms. Since publishing that work, we have expanded the number of platforms analyzed, collected more recent data, and selected more metrics to analyze. 

We have organized the rest of the paper as follows. In section~\ref{sec:Background} we provide some background information on blockchain, blockchain development platforms, and some of the technical aspects that differentiate blockchains. In section~\ref{sec:SelectedProjects} we present our 23 selected platforms and our methods for choosing those platforms. In section~\ref{sec:AnalysisOfBlockchainProjects} we analyze the selected blockchain development platforms on various metrics. Our analysis is broken into six subsections and covers the data presented in our seven tables. In section~ \ref{sec:Discussion}, we discuss how developers can use our analysis to select a blockchain development platform for their projects. Lastly, we provide some concluding thoughts in section~\ref{sec:Conclusion}.

\section{Background}
\label{sec:Background}

In its most basic form, a blockchain is a digital ledger that is tamper-resistant, tamper-evident, and often implemented in a distributed fashion. A blockchain maintains these properties through the use of cryptographic hashes. A blockchain is segmented into "blocks" that contain data. Each block contains the cryptographic hash of the previous block. Including the previous block's hash in a block creates a chain of blocks linked by hashes. That is, if the data is changed in any one block, the hashes must be updated in all subsequent blocks. For a more detailed description of blockchain see \cite{NISTBlockchain}, \cite{guoBlockchainTechnologySurvey_2022},and \cite{General_Blockchain_Description_2017}.

In this work, we do not look at specific blockchains but rather blockchain development platforms. A blockchain development platform is any platform that allows users to develop blockchain-based applications. For example, Ethereum is a blockchain development platform, whereas the Ethereum mainnet or the Ropsten testnet are examples of Ethereum blockchains. While this distinction may seem nuanced, it is an important distinction to make concerning permissioned blockchains. Platforms such as Hyperledger Fabric let developers build their own blockchains. In this case, it is clear that the blockchain development platform is different from the blockchain.

One way blockchains differentiate themselves is through the Sybil control mechanism~\cite{SybilControl}. Many previous works refer to this as the consensus protocol or the consensus algorithm. However, these mechanisms specifically prevent Sybil attacks; thus are properly called Sybil control mechanisms. In contrast, the consensus protocol is how a network agrees when a dispute, such as a fork in the blockchain. In Bitcoin, the Sybil control mechanism is Proof of Work (PoW), and the consensus protocol is that the network selects the longest chain when there is a fork. 

There are many different Sybil control mechanisms, and it is out of the scope of this work to discuss them here. However readers are encouraged to see \cite{ConsensusSurvey_2017}, \cite{ConsensusSurvey_2018}, and \cite{ConsensusSurvey_2020} for detailed surveys covering popular Sybil control mechanisms.

Another differentiators of blockchains is the permission structure. A blockchain can be either public or private, as well as permissionless or permissioned. These properties define how users can interact with a blockchain network. 

A public blockchain is a blockchain where any user can join and read the data on the blockchain. Likewise, a blockchain can be permissionless, where any user can join the Sybil control mechanism and attempt to add new blocks. In contrast, a private blockchain only allows a small group of users to read the data on the blockchain. Similarly, in a permissioned blockchain, users must be permitted to join the Sybil control mechanism and attempt to add new blocks. 

Lastly, blockchain development platforms can be categorized by their generation~\cite{TavaresSurveyOnBlockchainResearch2019}. Bitcoin, being the first blockchain, is a generation 1.X blockchain. This generation of blockchain focuses on a single use case for the blockchain and does not allow for much expansion. The Bitcoin blockchain only focuses on recording \$XBT transactions and does not allow developers to write expansive dapps. 

Wanting to expand blockchain technology further than just cryptocurrency, developers created generation 2.X blockchain development platforms such as Ethereum. These platforms allowed developers to write smart contracts. This generation of blockchain, for the first time, allowed developers to use blockchain in a variety of use cases and led to developers creating dapps and Web3. 

While generation 2.X development platforms are still popular with developers, generation 3.X platforms aim to create blockchains suited for enterprise applications. Some key features of generation 3.X blockchain development platforms include Layer 2 protocols, machine-to-machine (M2M) communication, mobile compatibility, use of DAGs, and the use of compute protocols.

In the following sections we will use the terminology discussed here to highlight features of various blockchain development platforms.

\section{Selected Projects}
\label{sec:SelectedProjects}

There are over 146000 blockchain projects on GitHub. Most of the projects were created only in the last few years. Since there are so many projects, it can be difficult for new developers to find a development platform that best suits their project's goals. This work aims to provide developers with the information necessary to select a platform to develop blockchain applications. Since there are so many blockchain development platforms, with many new ones being created each year, this work only analyzed popular open-source platforms that allow for dynamic smart contract development.  

We consider a project sufficiently popular if it has more than one-thousand stars on GitHub. The popularity restraint ensures that our analyzed blockchain development platforms are of high quality. We set our popularity requirement low enough to include many platforms yet high enough to ensure all of our discussed platforms are well known. However, popularity restraint favors older blockchains as they have had more time to amass a following. 

Likewise, we only analyzed open-source platforms so developers could see the underlying code. The added transparency of open source projects allows for more trusted code. 

Lastly, we only looked at platforms that allowed for dynamic smart contract development since our goal is to asst developers in creating new distributed applications. This restraint limits the projects analyzed to only platforms that allow developers to use generation 2.X or 3.X blockchains. 

In total, we considered 65 platforms; however, only 23 of those platforms meet our requirements. Some well-known platforms that did not satisfy our requirements include Cardano~\cite{CardanoWhitepaper}, Avalanche~\cite{AvalancheWhitepaper}, IOTA~\cite{IOTAWhitepaper}, and Lisk~\cite{LiskWhitepaper}. 

Table~\ref{sec:SelectedProjects} shows the projects we selected. We marked projects that are forks of older projects with an asterisk. Quorum and BNBChain are forks of Etherium, and Qtum is a fork of Bitcoin. Note that the data for all of our tables were collected in September 2022. Many whitepapers defining blockchain development platforms have not been submitted to peer-reviewed publications. Instead, many platforms publish their whitepaper directly to their website or GitHub page.

\begin{table}
\resizebox{\textwidth}{!}{%
\begin{tabular}{|l|l|l|l|}
\hline
\textbf{Platform} &
  \textbf{\begin{tabular}[c]{@{}l@{}}Github \\ Package Name\end{tabular}} &
  \textbf{\begin{tabular}[c]{@{}l@{}}Initial \\ Commit Date\end{tabular}} &
  \textbf{White Paper} \\ \hline
Ethereum                                                      & ethereum/go-ethereum                                                        & Dec 2013  & \cite{EthereumWhitepaper} \\ \hline
\begin{tabular}[c]{@{}l@{}}Hyperledger \\ Fabric\end{tabular} & hyperledger/fabric                                                          & May 2016  & \cite{HyperledgerFabricWhitepaper}, \cite{HyperledgerWhitepaper} \\ \hline
EOS                                                           & EOSIO/eos                                                                   & Apr 2017  &  \cite{EOSIOWhitepaper} \\ \hline
Solana                                                        & solana-labs/solana                                                          & Feb 2018  &  \cite{SolanaWhitepaper} \\ \hline
Tendermint                                                    & Tendermint/Tendermint                                                       & Apr 2014  &  \cite{TendermintWhitepaper} \\ \hline
Quorum                                                        & ConsenSys/quorum                                                            & Dec 2013* &  \cite{QuorumWhitepaper} \\ \hline
Corda                                                         & corda/corda                                                                 & Nov 2015  &  \cite{CordaWhitepaperPlatform},\cite{CordaWhitepaperTechnical} \\ \hline
Neo                                                           & neo-project/neo                                                             & May 2015  &  \cite{NeoWhitepaper} \\ \hline
Tron                                                          & tronprotocol/java-tron                                                      & Dec 2017  &  \cite{TronWhitepaper} \\ \hline
Stellar                                                       & stellar/stellar-core                                                        & Nov 2014  &  \cite{StellarWhitepaper} \\ \hline
Stacks                                                        & \begin{tabular}[c]{@{}l@{}}stacks-network/\\ stacks-blockchain\end{tabular} & Jan 2014  &  \cite{StacksWhitepaper} \\ \hline
BNB Chain                                                     & bnb-chain/bsc                                                               & Dec 2013* &  \cite{BNBWhitepaper} \\ \hline
\begin{tabular}[c]{@{}l@{}}Hyperledger \\ Sawtooth\end{tabular} &
  \begin{tabular}[c]{@{}l@{}}hyperledger/\\ sawtooth-core\end{tabular} &
  Mar 2016 &
   \cite{SawtoothWhitepaper}, \cite{HyperledgerWhitepaper} \\ \hline
NEAR                                                          & NEAR/nearcore                                                               & Oct 2018  &  \cite{NEARWhitepaper} \\ \hline
Tezos                                                         & tezos/tezos                                                                 & Sep 2016  &  \cite{TezosWhitepaper} \\ \hline
Optimism                                                      & \begin{tabular}[c]{@{}l@{}}ethereum-optimism/\\ optimism\end{tabular}       & Sep 2020  &  \cite{OptimismWhitepaper} \\ \hline
IoTeX                                                         & iotexproject/iotex-core                                                     & Apr 2018  &  \cite{IoTeXWhitepaper} \\ \hline
Harmony                                                       & harmony-one/harmony                                                         & May 2018  &  \cite{HarmonyWhitepaper} \\ \hline
Waves                                                         & wavesplatform/Waves                                                         & Jan 2015  &  \cite{WavesWhitepaper} \\ \hline
Algorand                                                      & algorand/go-algorand                                                        & Jun 2019  &  \cite{AlgorandWhitepaper} \\ \hline
Qtum                                                          & qtumproject/qtum                                                            & Aug 2009* &  \cite{QtumWhitepaper} \\ \hline
Exonum                                                        & exonum/exonum                                                               & Apr 2016  &  \cite{ExonumWhitepaper} \\ \hline
Zilliqa                                                       & Zilliqa/Zilliqa                                                             & Dec 2017  &  \cite{ZilliqaWhitepaper} \\ \hline
\end{tabular}%
}
\caption{We considered 65 platforms and selected the 23 most popular platforms for our analysis. Dates marked with a * represent packages that are forks of older packages.}
\label{tab:SelectedProjects}
\end{table}

In the following section, we will analyze our 23 selected platforms and provide metrics to assist developers in selecting a platform that best complements their projects. 

\section{Analysis of Blockchain Projects}
\label{sec:AnalysisOfBlockchainProjects}

\subsection{Popularity}
\label{sec:Popularity}

In the first stage of our analysis, we compared the popularity of our selected projects shown in Table~\ref{tab:Popularity}. Because of our popularity restraint, all 23 selected platforms have at least 1000 stars on GitHub. Thus, even the least popular of our selected platforms are well known. Developers can use our popularity measures to gauge how large the community surrounding a blockchain development platform is and how many developers are using a given platform. 

\begin{table}
\resizebox{1.25\textwidth}{!}{%
\begin{tabular}{|l|l|l|l|l|l|l|l|}
\hline
\textbf{Platform} &
  \textbf{\begin{tabular}[c]{@{}l@{}}Github \\ Forks\end{tabular}} &
  \textbf{\begin{tabular}[c]{@{}l@{}}Github \\ Stars\end{tabular}} &
  \textbf{\begin{tabular}[c]{@{}l@{}}Closed \\ Github Issues\end{tabular}} &
  \textbf{\begin{tabular}[c]{@{}l@{}}Stack\\ Exchange\end{tabular}} &
  \textbf{\begin{tabular}[c]{@{}l@{}}Google \\ Scholar\end{tabular}} &
  \textbf{Dapps} &
  \textbf{\begin{tabular}[c]{@{}l@{}}Popularity \\ Score\end{tabular}} \\ \hline
Ethereum                                                        & 14840 & 39163 & 6391 & 6317 & 18500 & 3452 & 100.00 \\ \hline
\begin{tabular}[c]{@{}l@{}}Hyperledger \\ Fabric\end{tabular}   & 8190  & 13950 & 162  & 1502 & 2310  & -    & 32.93  \\ \hline
EOS                                                             & 3700  & 11346 & 4751 & 81   & 275   & 579  & 19.54  \\ \hline
Solana                                                          & 2329  & 9372  & 3491 & 383  & 78    & 105  & 15.43  \\ \hline
Tendermint                                                      & 1786  & 5142  & 2800 & 68   & 138   & -    & 9.05   \\ \hline
Quorum                                                          & 1184  & 4234  & 757  & 206  & 379   & -    & 7.62   \\ \hline
Corda                                                           & 1069  & 3878  & 579  & 500  & 225   & -    & 7.20   \\ \hline
Neo                                                             & 1000  & 3381  & 1020 & 32   & 194   & 1    & 5.84   \\ \hline
Tron                                                            & 1139  & 3132  & 1030 & 121  & 43    & 1397 & 5.63   \\ \hline
Stellar                                                         & 992   & 3000  & 1157 & 33   & 190   & -    & 5.35   \\ \hline
Stacks                                                          & 545   & 2688  & 1749 & 219  & 20    & -    & 4.40   \\ \hline
BNB Chain                                                       & 788   & 1685  & 652  & 86   & 0     & 4067 & 3.25   \\ \hline
\begin{tabular}[c]{@{}l@{}}Hyperledger \\ Sawtooth\end{tabular} & 762   & 1393  & 4    & 113  & 135   & -    & 3.05   \\ \hline
NEAR                                                            & 364   & 1822  & 2064 & 170  & 13    & 43   & 3.01   \\ \hline
Tezos                                                           & 209   & 1501  & 0    & 46   & 212   & 80   & 2.50   \\ \hline
Optimism                                                        & 450   & 1397  & 434  & 4    & 4     & 25   & 2.35   \\ \hline
iotex                                                           & 291   & 1431  & 987  & 1    & 18    & 45   & 2.21   \\ \hline
Harmony                                                         & 268   & 1447  & 971  & 12   & 12    & 117  & 2.21   \\ \hline
Waves                                                           & 419   & 1164  & 343  & 67   & 72    & 16   & 2.18   \\ \hline
Algorand                                                        & 346   & 1132  & 5    & 21   & 112   & 12   & 2.04   \\ \hline
Qtum                                                            & 389   & 1162  & 292  & 11   & 28    & -    & 2.02   \\ \hline
Exonum                                                          & 245   & 1176  & 340  & 5    & 57    & -    & 1.88   \\ \hline
Zilliqa                                                         & 263   & 1110  & 264  & 1    & 15    & -    & 1.76   \\ \hline
\end{tabular}%
}
\caption{This table shows various popularity metrics for our selected platforms. Popularity Score is the normalized sum of the Forks, Stars, Stack Exchange Questions, and Google Scholar Articles.}
\label{tab:Popularity}
\end{table}

Since all of our projects have GitHub repositories, we could compare the number of GitHub Stars. A Star in GitHub allows a developer to like a project and save it for later. Thus this metric reflects how many developers are interested in each blockchain development platform. Likewise, we looked at the GitHub forks. The number of GitHub forks lets us know how many developers are interested in creating projects based on a given blockchain development platform.

Similarly, we examined the number of closed GitHub issues to see how many people were asking questions about the blockchain development platform. However, some platforms delete old issues when a new update to the blockchain development platform is released. Deleted issues explain why platforms like Hyperledger Fabric have fewer closed GitHub issues than expected. Because of the deleted issues, we do not include this metric when calculating popularity scores.

We also measured the popularity of the projects outside of GitHub using Stack Exchange. Stack Exchange is a website where users can ask questions and receive responses from other users with expertise in the platform. For all of our platforms, we searched Stack Exchange using the blockchain tag followed by the platform's name to eliminate questions unrelated to the blockchain. For example, when searching Waves in Stack Exchange, we revived many answers related to physics and not blockchain; thus, we searched for "[blockchain] Waves." While this may have removed some questions about each platform that did not include the blockchain tag, it ensured that all the questions returned were about the relevant blockchain development platform. 

Likewise, using Google Scholar, we measured the number of academic papers related to each blockchain development platform. Again for all of our selected platforms, we needed to restrict the results to only papers containing the platform followed by the word blockchain. For example, when searching for Optimism Blockchain returns many papers where the authors state their optimism about blockchain technology; thus, instead, we searched only for papers that included Optimism followed by blockchain, which returned only papers about the platform. Again this likely dropped some papers related to the platform, but it ensured that all the papers found were relevant to the platform. 

BNB Chain did not have any Google Scholar results. The lack of papers is because the blockchain was formally called the Binance Smart Chain (BSC). However, Binance merged the BSC with other blockchain technologies they were hosting and renamed it BNB Chain. We did not search for papers related to Binance Smart Chain because there are significant differences between BNB Chain and its predecessor. 

Lastly, we looked at how many dapps each blockchain development platform had posted to DappRadar. DappRadar is a website that aggregates and ranks known dApps. Naturally, none of the platforms that create private blockchains have any entries in DappRadar since the applications create on these platforms are not public. However, some public blockchain platforms, such as Stellar, have no entries in DappRadar. Since DappRadar does not collect information on all our selected projects, we did not include the number of dApps when calculating the popularity score. 

To calculate the popularity score, we took the normalized sums of GitHub forks, GitHub stars, StackExchange questions, and Google Scholar papers. The most popular platform can have a score of 100. 
	\[PopularityScore_{i}=\frac{forks_i+stars_i+questions_i+papers_i}{forks_{max}+stars_{max}+questions_{max}+papers_{max}}*100\]

Ethereum is the most popular platform, with a score of 100. Hyperledger Fabric is the most popular private blockchain development platform and the second most popular platform, scoring 32.93. While our least popular project Zilliqa only has a score of 1.76. The popularity score only allows us to compare the selected project to each other. All of the platforms shown have at least 1000 stars on GitHub, meaning they are all well known. 

A popular platform is not the best platform for every project. Developers should consider the popularity of a project only as a means to see how much support is available. We hope that by using this table in conjunction with our other tables, developers can choose the platform that best supports their projects.

\subsection{Properties of Blockchain Development Platforms}
\label{sec:BlockchainFundamentals}

Our subsequent table, Table~\ref{tab:fundamentals} describes the fundamental differentiators between blockchains tied to the blockchain development platforms. The table includes the blockchain generation, whether the blockchain is permissioned or permissionless, what Sybil control mechanism the blockchain uses if developers need to learn a framework-specific programming language to develop smart contracts, and if a cryptocurrency governs the blockchain. Note that some of our selected platforms are tied to a well-established mainnet, such as Ethereum, whereas other platforms will have developers create their own blockchain, such as Hyperledger Fabric.

\begin{table}
\resizebox{\textwidth}{!}{%
\begin{tabular}{|l|l|l|l|l|l|}
\hline
\textbf{Platform} &
  \textbf{Generation} &
  \textbf{Permissionless} &
  \textbf{Consensus} &
  \textbf{\begin{tabular}[c]{@{}l@{}}Prog. \\ Lang.\end{tabular}} &
  \textbf{\begin{tabular}[c]{@{}l@{}}Crypto-\\ currency\end{tabular}} \\ \hline
Ethereum &
  2.0 &
  Yes &
  PoS &
  Yes &
  \begin{tabular}[c]{@{}l@{}}\$ETH \\ (Ether)\end{tabular} \\ \hline
\begin{tabular}[c]{@{}l@{}}Hyperledger \\ Fabric\end{tabular} &
  2.x &
  No &
  \begin{tabular}[c]{@{}l@{}}Kafka, Solo, \\ BFT-SMaRt\end{tabular} &
  No &
  - \\ \hline
EOS &
  2.x &
  Yes &
  aBFT + DPoS &
  No &
  \begin{tabular}[c]{@{}l@{}}\$EOS \\ (EOSIO)\end{tabular} \\ \hline
Solana &
  2.x &
  Yes &
  \begin{tabular}[c]{@{}l@{}}PoH \\ (Similar to PoS)\end{tabular} &
  No &
  \begin{tabular}[c]{@{}l@{}}\$SOL \\ (Solana)\end{tabular} \\ \hline
Tendermint &
  2.x &
  No &
  pBFT &
  No &
  - \\ \hline
Quorum &
  2.x &
  No &
  PoA &
  Yes &
  - \\ \hline
Corda &
  2.x &
  No &
  Notary Nodes &
  No &
  - \\ \hline
Neo &
  3.x &
  Yes &
  \begin{tabular}[c]{@{}l@{}}dBFT \\ (Similar to PoS)\end{tabular} &
  No &
  \begin{tabular}[c]{@{}l@{}}\$NEO \\ (Neo) \\ \$GAS \\ (NeoGas)\end{tabular} \\ \hline
Tron &
  2.x &
  Yes &
  DPoS &
  Yes &
  \begin{tabular}[c]{@{}l@{}}\$TRX \\ (TRON)\end{tabular} \\ \hline
Stellar &
  2.x &
  Yes &
  FBA (Stellar) &
  No &
  \begin{tabular}[c]{@{}l@{}}\$XML \\ (Lumen)\end{tabular} \\ \hline
Stacks &
  2.x &
  Yes &
  \begin{tabular}[c]{@{}l@{}}PoX \\ (Proof of Transfer)\end{tabular} &
  Yes &
  \begin{tabular}[c]{@{}l@{}}\$STX \\ (Stacks)\end{tabular} \\ \hline
BNB Chain &
  2.x &
  Yes &
  pBFT \& PoSA &
  Yes &
  \begin{tabular}[c]{@{}l@{}}\$BNB \\ (BNB)\end{tabular} \\ \hline
\begin{tabular}[c]{@{}l@{}}Hyperledger \\ Sawtooth\end{tabular} &
  2.x &
  No &
  PoET, Raft &
  No &
  - \\ \hline
NEAR &
  2.x &
  Yes &
  Nightshade &
  No &
  \begin{tabular}[c]{@{}l@{}}\$NEAR\\ (NEAR)\end{tabular} \\ \hline
Tezos &
  2.x &
  Yes &
  PoS &
  No &
  \begin{tabular}[c]{@{}l@{}}\$XTZ \\ (Tezos)\end{tabular} \\ \hline
Optimism &
  3.x &
  Yes &
  DPoS &
  Yes &
  \begin{tabular}[c]{@{}l@{}}\$OP \\ (Optimism)\end{tabular} \\ \hline
iotex &
  3.x &
  Yes &
  DPoS &
  Yes &
  \begin{tabular}[c]{@{}l@{}}\$IOTX \\ (IoTeX)\end{tabular} \\ \hline
Harmony &
  3.x &
  Yes &
  pBFT &
  Yes &
  \begin{tabular}[c]{@{}l@{}}\$ONE \\ (Harmony)\end{tabular} \\ \hline
Waves &
  2.x &
  Yes &
  LPoS &
  Yes &
  \begin{tabular}[c]{@{}l@{}}\$WAVES \\ (Waves)\end{tabular} \\ \hline
Algorand &
  2.x &
  Yes &
  PPoS &
  No &
  \begin{tabular}[c]{@{}l@{}}\$ALGO \\ (Algorand)\end{tabular} \\ \hline
Qtum &
  2.x &
  No &
  MPoS &
  No &
  \begin{tabular}[c]{@{}l@{}}\$QTUM \\ (Qtum)\end{tabular} \\ \hline
Exonum &
  2.x &
  No &
  pBFT &
  No &
  - \\ \hline
Zilliqa &
  2.x &
  Yes &
  pBFT &
  Yes &
  \begin{tabular}[c]{@{}l@{}}\$ZIL \\ (Zilliqa)\end{tabular} \\ \hline
\end{tabular}%
}
\caption{This table shows key properties of our selected blockchain development platforms. The Prog. Lang. column represents if the project requires a framework-specific language, such as Solidity.}
\label{tab:fundamentals}
\end{table}

Most of our selected platforms are linked to generation 2.X blockchains. Of generation 3.X blockchains, both Optimism and Harmony are linked to layer 2 (L2) blockchains. An L2 blockchain performs transactions on a smaller side chain but uses a more extensive blockchain for finality. Both Optimism and Harmony finalize their transactions on the Ethereum mainnet blockchain. Developers should note that while L2s aim to increase transaction speed, they give up some security by using a smaller network. This trade-off of security for speed is essential in applications requiring exceptionally high network trust. However, the speed gained by using an L2 may outweigh the reduced security for many applications. 

Likewise, NEO and IoTeX allow for M2M communication, another property of generation 3.X blockchains. Developers wishing to create IoT applications should consider using these blockchain development platforms. 

Our selected projects contain a mix of permissioned and permissionless blockchains. Often enterprise or other private applications will require a permissioned blockchain. However, developers should be aware that when creating a premissioned blockchain, they may also need to spend time creating and operating blockchain nodes, creating an authority to grant permission to the blockchain network, or even choosing the consensus algorithm the network will use. For example, in Hyperledger Fabric, a developer needs to create their own private network before they can begin working on developing their dApp. While Hyperledger Fabric offers a suite of tools to make this step less complex, it still adds extra complications to the dApp development process. 

In contrast, public-facing applications will often require developers to use permissionless blockchains. While developers will not need their own networks, they will often need to pay a fee to use the public mainnet. Likewise, developers will have little control over the network; thus, a permissionless blockchain may not be suitable for applications handling sensitive information. Developers should consider which permission structure is best for their application. 

Some selected platforms, such as EOS, NEO, and NEAR, allow developers to use a permissionless network or create a private network. Access to two types of networks allows developers to switch which type of network they are using without switching platforms. For example, a developer may start off using a public mainnet. However, the developer may want to release a premium version of their application on a private blockchain. 

Our selected projects use a wide variety of Sybil control mechanisms, often called consensus algorithms. Even though the original blockchain, Bitcoin, uses Proof of Work (PoW) as its Sybil control mechanism, none of our selected projects use PoW. PoW has been shown to have many negative drawbacks compared to modern algorithms such as Proof of Stake (PoS), including high energy consumption~\cite{EnergyConsumption_PoW_PoS_2020} and slow performance~\cite{Performance_PoW_vs_PoS}. Thus PoS and algorithms similar to PoS are the most popular on our list. 

Developers should understand the pros and cons of the Sybil control mechanism. We suggest that developers read their selected project's whitepaper to understand how a specific blockchain development platform implements the Sybil control mechanism on their blockchains. 

Some blockchains, such as Hyperledger Fabric, allow the user to pick which mechanism the network will use when the user creates the network. This modular Sybil control mechanism layer may be desirable to developers who require specific control mechanisms for a given application. 

Before starting a project, developers should know if a blockchain requires a framework-specific programming language for developing smart contracts. The most well-known framework-specific programming language is Solidity, created initially to develop Ethereum dapps. Table~\ref{tab:smartcontractlanguge} shows all of the languages available for each project. Since smart contracts can be responsible for handling financial transactions, developers should make sure they understand the quirks of any language they choose. A small mistake in code can easily lead to a significant financial loss for the dApp's developers and users. 
  
Lastly, permissionless blockchains are governed by cryptocurrencies. Developers must pay a fee to run their applications on a permissionless blockchain. Some platforms have multiple cryptocurrencies. NEO has \$NEO for participating in PoS and \$GAS for paying the block creation fees. Since cryptocurrency prices are volatile, we do not report them here; however, developers should research the cost of creating dApps on their selected platform.

Developers should understand the fundamentals of each blockchain development platform before starting a project. Developers should ensure that the platform aligns with their project's goals. For example, a developer should select a permissionless blockchain if they plan to create an application that anyone can access. Understanding the fundamentals will aid developers in picking the best platform for their projects. 

\subsection{Existing Applications}
\label{sec:ExistingApplications}

Table~\ref{tab:applications} analyzes the various applications built using each blockchain development platform. Before starting an application, developers should be aware of the types of projects each platform is designed for. While many of these platforms are designed for general applications, some have particular use cases.

\begin{table}
\resizebox{1.25\textwidth}{!}{%
\begin{tabular}{|l|l|l|l|}
\hline
\textbf{Platform} &
  \textbf{Application Area} &
  \textbf{Example Use Cases} &
  \textbf{Sample DApps} \\ \hline
Ethereum &
  FinTech, general dApps &
  Decentalized Exchange, NFTs, Games &
  \begin{tabular}[c]{@{}l@{}}Uniswap, Crypto Kitties, \\ Open Sea\end{tabular} \\ \hline
\begin{tabular}[c]{@{}l@{}}Hyperledger \\ Fabric\end{tabular} &
  GovTec, Enterprise Applications &
  Voteing, B2B Supply Chains &
  BRUINchain, Healthchain \\ \hline
EOS &
  general dApps, Enterprise Applications &
  \begin{tabular}[c]{@{}l@{}}Bussiness Applications, Games, \\ Trading Digital Assets\end{tabular} &
  Defibox, Upland,  Everipedia \\ \hline
Solana &
  general dApps, FinTech &
  Gaming, DAO, DeFi Payment &
  \begin{tabular}[c]{@{}l@{}}Brave, Magic Eden, \\ Gameta\end{tabular} \\ \hline
Tendermint &
  FinTech &
  DEX, StableCoin, Carbon Trading &
  \begin{tabular}[c]{@{}l@{}}Biance DEX, Terra, \\ Regan Network\end{tabular} \\ \hline
Quorum &
  FinTech &
  \begin{tabular}[c]{@{}l@{}}Interbank Transfers, \\ Marketplace for Loans\end{tabular} &
  Project Ubin, Skeps \\ \hline
Corda &
  FinTech, Healthcare, Consturction &
  \begin{tabular}[c]{@{}l@{}}Capital Markets, Claims Managment, \\ Supply Chain\end{tabular} &
  HSBLOX \\ \hline
Neo &
  FinTech, IoT &
  Lottery, Social Network &
  Effect, Naritive \\ \hline
Tron &
  FinTech, general dApps &
  Decentalized Exchange, NFTs, Games &
  SunSwap, BSG, WINk \\ \hline
Stellar &
  FinTech &
  Interbank Transfers &
  IBM BWW \\ \hline
Stacks &
  FinTech, Bitcoin dApps &
  Decentralized Exchange,Domain Registrar &
  ALEX, Gamma, Megapont \\ \hline
BNB Chain &
  FinTech, general dApps &
  Decentalized Exchange, NFTs, Games &
  PancakeSwap, Era7, ApeSwap \\ \hline
\begin{tabular}[c]{@{}l@{}}Hyperledger \\ Sawtooth\end{tabular} &
  Healthcare, Enterprise Applictions &
  EHR, Enterprise Managment &
  PokitDok, Sextant \\ \hline
NEAR &
  FinTech, general dApps &
  Decentalized Exchange, NFTs, Games &
  Paras, Burrow, Mintbase \\ \hline
Tezos &
  FinTech, GovTech, general dApps &
  Stable Assets, Team Managment &
  lugh, Red Bull Racing, fxhash \\ \hline
Optimism &
  Fintech,general dApps &
  DeFi, Marketplaces, Exchanges &
  Quixotic,  Across, Rubicon \\ \hline
IoTeX &
  IoT &
  IoT Managment, IoT data sharing &
  Mimoswap, StarCrazy, VITA \\ \hline
Harmony &
  Fintec, general dApps &
  DeFi, NFTs, Decentralized Exchanges &
  Sushi, Timeless, Ript.io \\ \hline
Waves &
  general dApps &
  NFTs, Games, DeFi &
  \begin{tabular}[c]{@{}l@{}}Vires.finacnce, Hashgreed, \\ Waves Ducks\end{tabular} \\ \hline
Algorand &
  FinTech, general dApps &
  Decentalized Exchange, NFTs, Games &
  Algodex, Algofi, Octorand \\ \hline
Qtum &
  FinTech, Bitcoin dApps &
  Creating Tokens, Travel &
  QRC20 Tokens, Travala \\ \hline
Exonum &
  GovTech, dApps &
  Property Managment, eAuctions &
  \begin{tabular}[c]{@{}l@{}}Land Registry in Gerorgia, \\ Ukrain eAuction platform\end{tabular} \\ \hline
Zilliqa &
  FinTech, dApps &
  ePayments, Advertisement &
  Xfers, Aqilliz, Zilswap \\ \hline
\end{tabular}%
}
\caption{This table shows examples of existing applications for our selected blockchain development platforms.}
\label{tab:applications}
\end{table}

The most common application area of blockchain is in FinTech; however, blockchain has many use cases in other areas. Many private blockchain platforms, such as HyperLedger Fabric, focus on allowing users to create secure enterprise applications. Likewise, some blockchain development platforms, such as Exonum, have focused on allowing users to create government applications. Stellar is focused on interbank transfers, a particular FinTech use case. Developers must know which platforms are designed to support their projects sector.

Likewise, some blockchain development platforms focus on connectivity with existing blockchains. Optimism and Harmony are L2 blockchains; thus, they both focus on compatibility with the Ethereum mainnet blockchain. Similarly, Stacks and Qtum allow developers to create dApps that interact with the Bitcoin blockchain. Developers building applications that need to interact with these blockchains should consider what type of support within their chosen development platform.

To help developers best understand what types of applications they can build on each platform, we list some common use cases and some popular dApps for each project. Some unique projects include the F1 team Red Bull Racing using Tezos for team management and the country of Georgia using Exonum for land registry. While it is out of the scope of this work to discuss each of these dApps, developers are encouraged to research existing dApps. In particular, dApp code is available for any developer to read on public blockchains, which may help new developers understand common coding paradigms.  

Developers should familiarize themselves with some existing applications on each platform. This will help developers understand what applications are possible on each platform. For example, if a developer wishes to create an IoT application, they should consider platforms that are designed for IoT. The synergy between the platform and the project can make applications easier to build and more accessible to users.  

\subsection{Development Considerations}
\label{sec:DevelopmentConsiderations}

Table~\ref{tab:development} highlights key areas developers should understand before starting a project. The table covers the privacy features of each platform, the documentation available to developers, if there is a test network for developers, if there are tools available for developers, and how much control developers have over the Sybil control mechanism. 

\begin{table}
\resizebox{1.25\textwidth}{!}{%
\begin{tabular}{|l|l|l|l|l|l|}
\hline
\textbf{Platform} &
  \textbf{\begin{tabular}[c]{@{}l@{}}Privacy\\ Score\end{tabular}} &
  \textbf{\begin{tabular}[c]{@{}l@{}}Documentation \\ Score\end{tabular}} &
  \textbf{Test Network} &
  \textbf{\begin{tabular}[c]{@{}l@{}}Development\\  Tools\end{tabular}} &
  \textbf{\begin{tabular}[c]{@{}l@{}}Sybil \\ Modularity\end{tabular}} \\ \hline
Ethereum                                                        & 1 & 31 & Public         & Yes & None \\ \hline
\begin{tabular}[c]{@{}l@{}}Hyperledger \\ Fabric\end{tabular}   & 7 & 7  & Private        & Yes & High \\ \hline
EOS                                                             & 5 & 31 & Public         & Yes & Low  \\ \hline
Solana                                                          & 0 & 7  & Public         & Yes & None \\ \hline
Tendermint                                                      & 5 & 7  & Private        & Yes & None \\ \hline
Quorum                                                          & 7 & 7  & No             & Yes & None \\ \hline
Corda                                                           & 7 & 15 & Public         & No  & None \\ \hline
Neo                                                             & 4 & 3  & Public/Private & Yes & None \\ \hline
Tron                                                            & 5 & 3  & Public         & Yes & None \\ \hline
Stellar                                                         & 0 & 15 & Public         & Yes & None \\ \hline
Stacks                                                          & 0 & 3  & Public         & Yes & None \\ \hline
BNB Chain                                                       & 2 & 7  & Public         & Yes & None \\ \hline
\begin{tabular}[c]{@{}l@{}}Hyperledger \\ Sawtooth\end{tabular} & 5 & 3  & Public         & Yes & High \\ \hline
NEAR                                                            & 7 & 7  & Public/Private & No  & None \\ \hline
Tezos                                                           & 3 & 19 & No             & Yes & None \\ \hline
Optimism                                                        & 4 & 3  & Public/Private & Yes & None \\ \hline
IoTeX                                                           & 0 & 23 & Priavte        & No  & None \\ \hline
Harmony                                                         & 3 & 23 & Public         & Yes & None \\ \hline
Waves                                                           & 0 & 3  & No             & No  & None \\ \hline
Algorand                                                        & 0 & 23 & Private        & Yes & None \\ \hline
Qtum                                                            & 0 & 7  & Public         & Yes & None \\ \hline
Exonum                                                          & 5 & 3  & Private        & No  & None \\ \hline
Zilliqa                                                         & 0 & 3  & Public         & Yes & None \\ \hline
\end{tabular}%
}
\caption{This table highlights development features of our selected blockchain development platforms.}
\label{tab:development}
\end{table}

We assigned a privacy score to each platform. The privacy score lets developers know the privacy features available on each platform. Naturally, business applications may require more privacy than general applications such as games. Thus, developers must select a platform that will allow for the appropriate level of privacy. A project received 1 point if it allowed for private transactions, 2 points for Zero-Knowledge Proofs (ZKP), and 4 points if it allowed developers to create their own private blockchain. A ZKP is a way for two actors to interact with each other without giving up their identities. ZCash~\cite{ZCashWhitepaper} is a popular cryptocurrency that implements ZKP to allow for private blockchain transactions. For detailed information on ZKP, see~\cite{def_and_prop_ZKP_1994} and \cite{survey_ZKP_blockchain_2021}.

Since we assign points as powers of two, a privacy score can easily be decomposed into its parts by converting the score to binary. For example, Tron received a score of 5 points, which in base 2 is bin(101). Reading the binary from left to right, we see that it does allow for the creation of private blockchains (bin(100)), it does not have support for ZKP (bin(00)), and it does support private transactions (bin(1)).

Using a similar point system, we assign each project a documentation score. This metric aims to show developers how much support is available from the platform. A project was given 1 point for having official written documentation, 2 points for having a simple tutorial on how to create an application, 4 points if the documentation showed developers how to build a complete example application, 8 points if the documentation included videos, and 16 points if the platform provided extensive documentation. Some examples of extensive documentation include Ethereum and Tezos having games developers can play to learn the basics of the platform or EOS having an official training course developers can take. 

The next column notes if the platform has a test network. Like in Web2 development, developers often do not want to deploy their code to production until they are done testing it. A test network is a blockchain where developers can test their dApps. Like all other blockchains, a testnet can be either public or private. A private testnet will require developers to create their own environment for testing. 

Often a faucet will accompany a public test. A faucet will give users cryptocurrency to pay block creation fees on a public testnet. The cryptocurrency given by the faucet is different from the cryptocurrency for the mainnet. Likewise, faucets rely on community support and can become dry; this may lead to issues for developers hoping to test on public testnets. 

Next, we note which platforms have official development tools. Development tools are created by a blockchain development platform to assist developers in creating dApps. For example, Ethereum has the Remix IDE, an online IDE, and many frameworks for testing dApps. Developers should be aware of development tools available that can assist them in creating dApps. 

Lastly, we cover the modularity of each project's Sybil Control layer, often called the consensus layer. Developers may need to optimism the Sybil control mechanism to best fit their application. Few platforms allow developers the flexibility to change the Sybil control layer. Hyperledger Fabric and Sawtooth allow developers to select from a set of Sybil control mechanisms (Shown in Table~\ref{tab:fundamentals}). Likewise, developers can modify part of the Sybil control mechanism if they create a private EOS network. Notice that no other platforms, including the platforms for private blockchains, allow developers to modify the Sybil Control layer, making the Hyperledger platforms and EOS unique. 

Developers should consider the privacy features and Sybil control layers' flexibility to ensure their selected platform supports their project. Likewise, newer developers may consider platforms with plenty of documentation, an easy-to-use testnet, and development tools. Using our table, developers can best select a platform that fits their skill level and project goals. 

\subsection{Smart Contract Features}
\label{sec:SmartContractFeatures}

The following tables highlight the features of smart contracts provided by each platform. Table~\ref{tab:smartcontractlanguge} shows which languages developers can use to write smart contracts. Languages marked with a * are languages designed for smart contract development.

\begin{table}
\resizebox{1.25\textwidth}{!}{%
\begin{tabular}{|l|l|}
\hline
\textbf{Platform}                 & \textbf{Smart Contract Development Language}                                \\ \hline
Ethereum                          & Solidity*, Viper*                                                          \\ \hline
Hyperledger Fabric                & Go, Java, Javascript, Typescript                                           \\ \hline
EOS                               & C++                                                                        \\ \hline
Solana                            & Rust, C++, C                                                               \\ \hline
Tendermint                        & Go, Python, Cosmos CLI*                                                    \\ \hline
Quorum                            & Solidity*                                                                  \\ \hline
Corda                             & Kotlin, Java                                                               \\ \hline
Neo                               & Python, C\#, Go, Java, Javascript, Typescript                              \\ \hline
Tron                              & Solidity*                                                                  \\ \hline
Stellar                           & Javascript, Java, Go, Python, C\#.NET, Ruby, IOS, Scala, Qt/C++, Flutter   \\ \hline
Stacks                            & Clarity*                                                                   \\ \hline
BNB Chain 												& Solidity*                                                                  \\ \hline
NEAR                              & Javascript, Rust, NEAR API                                                 \\ \hline
Hyperledger Sawtooth              & Python, Go, Javascript, Rust, Java, C++, Swift                             \\ \hline
Optimism                          & Solidity*                                                                  \\ \hline
Tezos                             & Python, OCaml, Javascript, Pascal, Reason, Indigo,  Archtype, Michleson* \\ \hline
IoTeX                             & Solidity* , Javascript, Java, Go, Swift, C                                 \\ \hline
Harmony                           & Solitdity*                                                                 \\ \hline
Waves                             & Ride*                                                                      \\ \hline
Qtum                              & Javascript, Solidity*                                                      \\ \hline
Algorand                          & Python, Reach                                                             \\ \hline
Exonum                            & Rust, Java                                                                 \\ \hline
Zilliqa                           & Scilla*                                                                    \\ \hline
\end{tabular}%
}
\caption{This table shows which programming languages can be used to develop smart contracts for each platform. Languages marked with a * are blockchain specific languages.}
\label{tab:smartcontractlanguge}
\end{table}

Our list of platforms contains a variety of languages, with the most common languages being Javascript and Solidity. Very few platforms allow for mobile smart contracts. Only Stellar and Hyperledger Sawtooth allow developers to use mobile languages like Swift. 

Framework-specific languages, such as Solidity, offer the benefit of being designed for blockchain development at the cost of having developers learn a new programming language. When selecting a platform, developers should recognize the pros and cons of using a framework-specific language. For example, take a developer who wants to create a financial application. A small error in the code could lead to a significant financial loss. On the one hand, the developer may want to program in a language they already know since they already understand the nuances of the language, but they may have a blockchain-related bug in their code. On the other hand, they may want to choose a framework-specific language since it will help the developer avoid blockchain errors. Still, the developer must learn the language and understand its nuances.  

Table~\ref{tab:smartContractsFeatures} shows which platforms are directly upgradable after deployment. Since smart contracts are written on a blockchain, which is immutable, developers can never change the original code after being published. However, some platforms have systems to ensure that developers can update their code, and the old code is ignored. While developers have been able to replicate these systems on most platforms, only a few of our selected platforms allow developers to update their code after deployment directly. 

\begin{table}
\resizebox{1.25\textwidth}{!}{%
\begin{tabular}{|l|l|l|l|l|l|}
\hline
\textbf{Platform} &
  \textbf{\begin{tabular}[c]{@{}l@{}}Directly \\ Upgradable\end{tabular}} &
  \textbf{\begin{tabular}[c]{@{}l@{}}Ricardian \\ Contracts\end{tabular}} &
  \textbf{\begin{tabular}[c]{@{}l@{}}Execution \\ Model\end{tabular}} &
  \textbf{OS Model} &
  \textbf{Compiled Languge} \\ \hline
Ethereum &
  No &
  No &
  \begin{tabular}[c]{@{}l@{}}Order-Execute \\ (OE)\end{tabular} &
  \begin{tabular}[c]{@{}l@{}}Ethereum \\ Virtual Machine\end{tabular} &
  Ethereum Bytecode \\ \hline
\begin{tabular}[c]{@{}l@{}}Hyperledger \\ Fabric\end{tabular} &
  Yes &
  Yes &
  \begin{tabular}[c]{@{}l@{}}Execute-Order-\\ Validate (EOV)\end{tabular} &
  Hyperledger OS &
  - \\ \hline
EOS &
  Yes &
  Yes &
  OE &
  \begin{tabular}[c]{@{}l@{}}EOSIO \\ Virtual Machine\end{tabular} &
  WebAssembly \\ \hline
Solana &
  Yes &
  No &
  \begin{tabular}[c]{@{}l@{}}OE: \\ Parallel Execution\end{tabular} &
  \begin{tabular}[c]{@{}l@{}}Solana \\ Virtual Machine\end{tabular} &
  Solana Bytecode \\ \hline
Tendermint &
  No &
  No &
  OE &
  ABCI Protocol &
  ABCI \\ \hline
Quorum &
  No &
  No &
  OE &
  \begin{tabular}[c]{@{}l@{}}Ethereum \\ Virtual Machine\end{tabular} &
  Ethereum Bytecode \\ \hline
Corda &
  Yes &
  Yes &
  \begin{tabular}[c]{@{}l@{}}If Then Else \\ (ITE)\end{tabular} &
  Corda API &
  REST API \\ \hline
Neo &
  Yes &
  No &
  OE &
  Neo Virtual Machine &
  Neo Bytecode \\ \hline
Tron &
  No &
  No &
  OE &
  Tron Virtual Machine &
  Ethereum Bytecode \\ \hline
Stellar &
  No &
  No &
  ITE &
  Stellar API &
  REST API \\ \hline
Stacks &
  No &
  No &
  EOV &
  Clarity Interpreter &
  - \\ \hline
BNB Chain &
  No &
  No &
  OE &
  \begin{tabular}[c]{@{}l@{}}Ethereum \\ Virtual Machine\end{tabular} &
  Ethereum Bytecode \\ \hline
\begin{tabular}[c]{@{}l@{}}Hyperledger \\ Sawtooth\end{tabular} &
  Yes &
  No &
  OE &
  \begin{tabular}[c]{@{}l@{}}Sawtooth \\ State Machines\end{tabular} &
  WebAssembly \\ \hline
NEAR &
  No &
  No &
  OE &
  NEAR Virtual Machine &
  WebAssembly \\ \hline
Tezos &
  No &
  No &
  \begin{tabular}[c]{@{}l@{}}OE: \\ Functional Progarming\end{tabular} &
  \begin{tabular}[c]{@{}l@{}}Michelson \\ Functional Programming\end{tabular} &
  Michelson \\ \hline
Optimism &
  No &
  No &
  OE &
  \begin{tabular}[c]{@{}l@{}}Optimism \\ Virtual Machine\end{tabular} &
  Ethereum Bytecode \\ \hline
IoTeX &
  No &
  No &
  OE &
  \begin{tabular}[c]{@{}l@{}}Ethereum \\ Virtual Machine\end{tabular} &
  Ethereum Bytecode \\ \hline
Harmony &
  No &
  No &
  OE &
  \begin{tabular}[c]{@{}l@{}}Ethereum \\ Virtual Machine\end{tabular} &
  Ethereum Bytecode \\ \hline
Waves &
  No &
  No &
  ITE &
  Waves API &
  REST API \\ \hline
Algorand &
  Yes &
  No &
  OE &
  \begin{tabular}[c]{@{}l@{}}Algorand \\ Virtual Machine\end{tabular} &
  TEAL \\ \hline
Qtum &
  No &
  No &
  OE &
  x86 Virtual Machine &
  x86 Assembly \\ \hline
Exonum &
  Yes &
  No &
  OE &
  Java Virtual Machine &
  Java Bytecode \\ \hline
Zilliqa &
  No &
  No &
  EOV &
  Scilla Interpreter &
  - \\ \hline
\end{tabular}%
}
\caption{This table highlights features of smart contracts on our selected blockchain development platforms.}
\label{tab:smartContractsFeatures}
\end{table}

Some of our selected platforms allow developers to create Ricardian Contracts. While smart contracts are often compared to legal agreements, they are not legal contracts. A Ricardian Contract is a legal contract that machines and people can read~\cite{RicardianContract}. Developers wishing to create applications requiring legal contracts should consider a platform that allows for Ricardian Contracts. 

Lastly, we highlight the implementation of smart contracts on each platform. The execution model shows when transactions are ordered and when the smart contract is executed. For example, in the Order-Execute model, transactions are first ordered, then the code is executed. In contrast, in the Execute-Order-Validate model, transactions are executed, and the output of those transactions is then ordered into a block. The Hyperledger Fabric Whitepaper~\cite{HyperledgerFabricWhitepaper} goes into detail on each type of execution model. 

Likewise, we note the OS model behind each blockchain development platform. Most of our platforms run on a distributed virtual machine, such as the Ethereum Virtual Machine (EVM). Each project's whitepaper contains information on how its virtual machine is implemented. 

Lastly, we show the compiled language that each model uses. While developers do not need to be able to write code in the compiled language, it can be helpful to understand the entire model when debugging. In traditional applications, developers do not need to know how to write in assembly to create apps. However, understanding that the system will compile their program to assembly can make debugging easier for developers. The same idea applies to Web3 development; developers should understand how their applications are compiled so they can more easily debug their code. 

By understanding the languages allowed by each platform, developers can best select a platform whose language they are comfortable with. Likewise, developers should understand what features the smart contracts provide to ensure they select a platform that synergies with their project. Using this table, developers can understand how smart contracts work on each platform to select a platform that fits their projects' needs.

\subsection{Community Engagement}
\label{sec:CommunityEngagement}

Table~\ref{tab:community} summarizes the community around each blockchain development platform. Developers should be aware of size and engagment of a community, so they know where to turn when they need technical assistance. Likewise, since blockchain focuses on decentralization, the community is often the decision maker for substantial changes to the platform. 

\begin{table}
\resizebox{1.25\textwidth}{!}{%
\begin{tabular}{|l|l|l|l|l|l|l|l|}
\hline
\textbf{Platfrom} &
  \textbf{\begin{tabular}[c]{@{}l@{}}Supporters\end{tabular}} &
  \textbf{\begin{tabular}[c]{@{}l@{}}Last \\ Commit\end{tabular}} &
  \textbf{\begin{tabular}[c]{@{}l@{}}Total\\ Commits\end{tabular}} &
  \textbf{Road Map} &
  \textbf{\begin{tabular}[c]{@{}l@{}}Offical \\ Telegram/\\ Discord\end{tabular}} &
  \textbf{\begin{tabular}[c]{@{}l@{}}Youtube \\ Videos\end{tabular}} &
  \textbf{Subreddit} \\ \hline
Ethereum &
  \begin{tabular}[c]{@{}l@{}}Vitalik Buterin, \\ Etherium Foundation\end{tabular} &
  Sep 2022 &
  13637 &
  Yes &
  Yes &
  1094 &
  1444724 \\ \hline
\begin{tabular}[c]{@{}l@{}}Hyperledger \\ Fabric\end{tabular} &
  \begin{tabular}[c]{@{}l@{}}Linux Foundation, \\ IBM, Intel, SAP Arabia\end{tabular} &
  Sep 2022 &
  14130 &
  No &
  Yes &
  1169 &
  3550 \\ \hline
EOS &
  \begin{tabular}[c]{@{}l@{}}Block.one, \\ EOSIO Foundation\end{tabular} &
  July 2021 &
  20554 &
  Yes &
  Yes &
  103 &
  97768 \\ \hline
Solana &
  \begin{tabular}[c]{@{}l@{}}Anatoly Yakovenko, \\ Solana Foundation\end{tabular} &
  Sep 2022 &
  20203 &
  Yes &
  Yes &
  314 &
  152831 \\ \hline
Tendermint &
  Jae Kwon, Cosmos/Ignite &
  Sep 2022 &
  8956 &
  Yes &
  Yes &
  359 &
  274 \\ \hline
Quorum &
  ConsenSys, JP Morgan &
  Sep 2022 &
  14311 &
  No &
  No &
  83 &
  2131 \\ \hline
Corda &
  R3 &
  Sep 2022 &
  9732 &
  Yes &
  No &
  257 &
  469 \\ \hline
Neo &
  \begin{tabular}[c]{@{}l@{}}Da Hongfei, Erik Zhang, \\ Onchain\end{tabular} &
  Sep 2022 &
  1388 &
  Yes &
  No &
  148 &
  117601 \\ \hline
Tron &
  Justin Sun, TRON DAO &
  Sep 2022 &
  17087 &
  No &
  Yes &
  559 &
  124952 \\ \hline
Stellar &
  \begin{tabular}[c]{@{}l@{}}Jed McCaleb, Joyce Kim, \\ Stellar Development Foundation\end{tabular} &
  Sep 2022 &
  7990 &
  Yes &
  Yes &
  143 &
  212769 \\ \hline
Stacks &
  \begin{tabular}[c]{@{}l@{}}Princeton, Hiro, \\ Stacks Foundation\end{tabular} &
  Sep 2022 &
  16993 &
  No &
  Yes &
  494 &
  6652 \\ \hline
BNB Chain &
  Binance &
  Sep 2022 &
  13506 &
  Yes &
  Yes &
  135 &
  17192 \\ \hline
\begin{tabular}[c]{@{}l@{}}Hyperledger \\ Sawtooth\end{tabular} &
  \begin{tabular}[c]{@{}l@{}}Linux Foundation, \\ IBM, Intel, SAP Arabia\end{tabular} &
  Jul 2022 &
  4590 &
  No &
  Yes &
  1169 &
  3550 \\ \hline
NEAR &
  \begin{tabular}[c]{@{}l@{}}Illia Polosukhin, Alexander Skidanov, \\ NEAR Foundation\end{tabular} &
  Sep 2022 &
  8251 &
  Yes &
  Yes &
  378 &
  12254 \\ \hline
Tezos &
  \begin{tabular}[c]{@{}l@{}}Dynamic Ledger Solutions, \\ Tezos Foundation\end{tabular} &
  Sep 2022 &
  3779 &
  No &
  Yes &
  171 &
  71141 \\ \hline
Optimism &
  Optimism Foundation &
  Sep 2022 &
  19477 &
  Yes &
  Yes &
  17 &
  5 \\ \hline
IoTeX &
  IoTeX &
  Sep 2022 &
  2831 &
  Yes &
  Yes &
  360 &
  18044 \\ \hline
Harmony &
  \begin{tabular}[c]{@{}l@{}}Stephen Tse, \\ Harmony Foundation\end{tabular} &
  Sep 2022 &
  7590 &
  Yes &
  Yes &
  406 &
  53477 \\ \hline
Waves &
  Waves Technology &
  Sep 2022 &
  12971 &
  No &
  Yes &
  86 &
  1484 \\ \hline
Algorand &
  Algorand &
  Sep 2022 &
  34919 &
  No &
  Yes &
  183 &
  75645 \\ \hline
Qtum &
  Qtum &
  Jun 2022 &
  3270 &
  Yes &
  Yes &
  98 &
  17679 \\ \hline
Exonum &
  Bitfury &
  Sep 2022 &
  5652 &
  Yes &
  Yes &
  7 &
  46 \\ \hline
Zilliqa &
  \begin{tabular}[c]{@{}l@{}}National University of Singapore, \\ Zilliqa Foundation\end{tabular} &
  Sep 2022 &
  9383 &
  Yes &
  Yes &
  132 &
  44459 \\ \hline
\end{tabular}%
}
\caption{This table highlights the communities sourding our selected blockchain development platforms. The last commit data was collected in September 2022.}
\label{tab:community}
\end{table}

We first recorded who the supporters of the blockchain development platform are. Blockchain development platforms supported by companies will have different end goals than platforms supported by individuals or foundations. Developers should keep in mind the goals of the development platform backers.

Next, we record when each platform last made a commit to GitHub. We recorded our dates in September 2022. In addition, we recorded the total number of commits. These two metrics show how active a blockchain development platform community is. When we recorded our data, only EOS had not made a commit within the last year.

Similarly, we record if the platform has an official road map. The road map outlines future changes the platform plans to make and when it aims to make those changes. Developers should know how active a platform is and what updates are planned. Some applications will require no modifications to the platform after deployment, while others will benefit from continued upgrades. 

We measured how active each platform's community is on social media. We note which platforms have an official Telegram or Discord channel. These are popular messaging applications where developers can join a group and ask others in the community questions about the platform. Likewise, we recorded how many YouTube videos were posted by the platform. Lastly, we recorded the number of members on each platform's subreddit. A subreddit is a forum on the social media platform Reddit dedicated to a specific topic. The platforms tied to a cryptocurrency have significantly more subreddit members than platforms without a cryptocurrency. This disparity is so drastic that the least popular platform Zilliqa, linked to the \$ZIL cryptocurrency, has more subreddit members than our second most popular project Hyperledger Fabric, which has no cryptocurrency. 

Developers should use our community table to judge how active a platform's community is to understand how much support will be available. Projects with a larger community will have more people who can assist a new developer. Likewise, developers should understand how active the community is and who the stakeholders are so they know where the project is going in the future and when they can expect the new features to get implemented. 

\section{Discussion}
\label{sec:Discussion}

This section outlines some key questions developers should ask before selecting a blockchain development platform. In addition to our tables, we hope this guides developers in selecting a platform that best meets their abilities and project goals.

First, developers should ask themselves if they need a permissionless or a permissioned blockchain. If a developer plans on only having a few select users, for example, in a business application, then a permissioned platform may be more appropriate. A permissionless platform will better suit the project if a developer plans to allow anyone to interact with their project. Developers should ask themselves questions such as "Should anyone be allowed to interact with the smart contract?" or "Dose smart contract code need to be confidential?"' 

Similarly, developers should consider if they need a public mainnet. Developers should understand the fees and transaction speed if they use a public mainnet. In contrast, if the developer chooses to create their own private mainnet, they should understand the cost of creating and maintaining that network. If a business is tracking shipments, it will likely want to keep that data hidden from competitors; thus, the developer will need to create a private network. In contrast, if a developer is not storing sensitive information on the blockchain, a public network may be more accessible. Questions such as "Do I want to use an existing blockchain infrastructure?", "What is the maximum that any execution of my smart contract can cost?" are essential in making this decision. 

Next, developers must ensure that the platform allows them to create their projects. For example, a developer wishing to create a banking application should look to use a platform that others have used to create similar applications. The developers should consider a few platforms for any project since one platform might be better optimized for the developer's project. For example, a developer may wish to use Ethereum for a banking application. However, they should also consider platforms like Stellar since it is more optimized for this type of project. "Have similar applications been created using this blockchain development platform?" and "Which blockchain development platform specializes for my use case?" can help developers choose the best platform. 

Developers should ensure they are comfortable with the language smart contracts are developed. Knowing the nuances of the language are particularly important when creating financial applications, as a mistake in the code can cause financial harm. To quickly get started, developers may wish to write in languages they are already familiar with. However, if developers spend time learning a framework-specific language designed for smart contract development, it may make more complicated blockchain tasks easier to program. Developers should consider, "Do I know the language for smart contract development already?" "Am I willing to learn a new blockchain-specific language?"

Similarly, developers need to understand what types of development tools are available. The development tools can help developers create, test, and maintain smart contracts. Developers should understand "What tools exist to help developers?" and "What type of support is there for new developers?"

Lastly, developers should understand the size of the project and the amount of community engagement surrounding a project. A large active community can help developers with technical issues when creating their dapps. Developers should consider questions such as "How popular is the blockchain development platform?" or "Is the community large enough to assist new developers?"

\section{Conclusion}
\label{sec:Conclusion}

In this work, we first outline the basics of blockchain so that developers can understand what differentiates various blockchain development platforms. Next, we discuss selecting 23 blockchain development platforms from our more extensive list. Then, we analyzed seven tables highlighting the key metrics developers should know for 23 selected platforms. We finished with a brief discussion on how developers can effectively use our metrics when creating new projects. We hope developers use our metrics to select the best project platform. 

In future work, we aim to create the same sample project on all our selected platforms and post those projects on GitHub. This GitHub page will allow developers to look at code to understand the nuances of each project better. We hope that the addition of code will assist new developers in selecting a platform and quickly getting started on projects. 

\bibliographystyle{plain}
\bibliography{ref}

\end{document}